\documentclass{ws-ijmpa}
\usepackage[super,compress]{cite}
\usepackage{graphicx}
\usepackage{amsfonts,amscd,amssymb,amsmath,latexsym}

\newcommand{\FF}{\mathcal{F}}

\begin{document}
\markboth{I. Dimitrijevic}{Some exact anisotropic cosmological solutions of a simple nonlocal de Sitter gravity}

%
\catchline{}{}{}{}{}
%

\title{Some exact anisotropic cosmological solutions of a simple nonlocal de Sitter gravity
}

\author{Ivan Dimitrijevic
}

\address{University of Belgrade, Faculty of Mathematics, Studentski Trg 16\\
Belgrade, Serbia \\
ivan.dimitrijevic@matf.bg.ac.rs}

%

\maketitle

\begin{history}
\received{Day Month Year}
\revised{Day Month Year}
\end{history}

\begin{abstract}
It was shown recently that a very simple nonlocal de Sitter gravity model contains exact vacuum cosmological  solution which mimics dark energy and dark matter in flat space. Some other interesting solutions have been also found. In this paper we proceed with finding several new exact cosmological solutions  which belong to Bianchi I space. These solutions are simple generalizations of solutions previously found in the FLRW case of the same nonlocal de Sitter gravity model. Obtained results are discussed.

\keywords{Anisotropic Universe; Cosmological solutions; Nonlocal de Sitter gravity.}
\end{abstract}

\ccode{PACS numbers:04.50.Kd, 04.20.Jb, 02.40.Ky}


\section{Introduction}
Current state of the Universe is very well described by the Standard Model of Cosmology (SMC) which is mainly based on General Relativity (GR), the Standard Model of Particle Physics (SMPP), observation that the Universe is homogeneous and isotropic at the very large cosmic scales and has an accelerating expansion. According to the SMC, the Universe at present time  consists of about 68\% of dark energy (DE), 27\% of dark matter (DM) and only 5\% of standard matter described by the SMPP\cite{Planck2018}. DM was introduced as a possible explanation of large velocities within and between clusters of galaxies.  After discovery of the accelerating expansion of the Universe in 1998, it was introduced DE as a new kind of matter with negative pressure that acts as antigravity causing this acceleration. According to the SMC, DE is related to the cosmological constant $\Lambda$. Therefore, the SMC is also known as the $\Lambda$CDM, where CDM means cold dark matter.

Since existence of dark matter and dark energy has not yet been experimentally confirmed, some researchers turned to look for alternative explanation of flat rotation curves in spiral galaxies, as well as the late time cosmic acceleration\cite{faraoni,nojiri,clifton,nojiri1,capozziello}. Practically, it means modification of the geometric sector of general relativity. Since there is no, so far, a guiding principle how  to choose appropriate extension of the Einstein-Hilbert (EH) action, there are many phenomenological approaches. Usually, these approaches are some extensions of the scalar curvature $R$ in the EH action by various forms of scalars that can be constructed in the pseudo-Riemannian geometry. The most elaborated version has been $f(R)$ gravity\cite{faraoni}, where $R$ is replaced by a function  $f(R)$. One of the actual and attractive approaches to the extension of GR is nonlocal modified gravity, see review\cite{capozziello}. Note that general relativity, despite enormous success,  has its own problems like black hole and big bang singularity, and problems with its quantization\cite{modesto}.

Recently, it was shown that a very simple nonlocal de Sitter gravity model, given by action
\begin{align} \label{eq.1.1}
S = \frac{1}{16 \pi G} \int  \big(R- 2 \Lambda  + \sqrt{R-2\Lambda}\ {\FF}(\Box)\ \sqrt{R-2\Lambda}\big) \sqrt{-g}\ \mathrm d^4 x ,
\end{align}
 contains exact vacuum cosmological  solution which mimics dark energy and dark matter in flat space\cite{dimitrijevic10}. Some other interesting solutions have been also found\cite{dimitrijevic13,dimitrijevic20}. In \eqref{eq.1.1},  $R$ is scalar curvature, $\Lambda$ is cosmological constant and $\FF(\Box) = \sum_{n=1}^{+\infty} f_n \Box^n + \sum_{n=1}^{+\infty} f_{-n} \Box^{-n}$ is nonlocal operator with d'Alembertian $\Box$. In this paper, we present several exact vacuum anisotropic cosmological solutions of the Bianchi I type of the same nonlocal gravity model \eqref{eq.1.1}, which are connected with its exact solutions in the Friedmann-Lema\^itre-Robertson-Walker (FLRW) metric case. While at the very large cosmic scales the Universe is homogeneous and isotropic, nevertheless anisotropic solutions are interesting as exact solutions and may be of interest for the very early evolution of the Universe, e.g. see recent references in modified gravity\cite{nojiri2,devi}. It is worth mentioning  that in recent article\cite{kumar} an anisotropic bouncing cosmological solution in higher-derivative non-local gravity was found.

This paper is organized as follows. In Sec. 2 metric of the Bianchi I space is presented. Nonlocal gravity model \eqref{eq.1.1} and the corresponding equations of motion are considered in Sec. 3. Some anisotropic cosmological solutions are investigated in Sec. 4. Sec. 5 contains some concluding remarks.

\section{The metric}
Let us consider  the Bianchi type I anisotropic metric in the form
\begin{equation}\label{metric:1}
  \mathrm{d}s^2 = -\mathrm d t^2 + a_1(t)^{2}\mathrm dx^2 + a_2(t)^{2}\mathrm dy^2 + a_3(t)^{2}\mathrm dz^2,
\end{equation}
with three scale factors $a_1(t)$, $a_2(t)$ and $a_3(t)$. It is worth noting that if all three scale factors are equal one obtains flat $FLRW$ metric. Also d'Alembertian of the metric \eqref{metric:1} reads
\begin{equation}
  \Box u(t)= - \ddot u(t) - (H_1(t) + H_2(t) + H_3(t))\dot u(t),
\end{equation}
where $H_i(t) = \frac{\dot a_i(t)}{a_i(t)}$.

Therefore if we introduce the Hubble parameter $H(t)$ by
\begin{align}
  H(t) &= \frac 13 (H_1(t) + H_2(t) + H_3(t)),
\end{align}
we obtain the same d'Alembertian as in the $FLRW$ metric. By integrating $H(t)$ one obtains the corresponding scale factor
\begin{align}
  a(t) &= \sqrt[3]{a_1(t)a_2(t)a_3(t)}. \label{frw:a}
\end{align}
For further calculations in the sequel it would be convenient to introduce the following notation
\begin{equation}\label{ai}
  a_i(t) = a(t) e^{\beta_i(t)}, \; i=1,2,3.
\end{equation}

Moreover, we will consider functions $\beta_1$, $\beta_2$ and $\beta_3$ as a components of a curve $\beta(t) = (\beta_1(t), \beta_2(t), \beta_3(t))$ in $\mathbb{R}^3$. Using the condition \eqref{frw:a} it is easy to see that $\beta(t)$ is a plane curve that lies in a plane $x+y+z =0 $, i.e.
\begin{equation}
  \beta_1(t)+ \beta_2(t) + \beta_3(t) =0.
\end{equation}

Let us denote the speed of curve $\beta(t)$ by $\sigma(t)$, hence
\begin{equation}
  \sigma(t)^2 = \dot \beta_1(t)^2+ \dot \beta_2(t)^2 + \dot \beta_3(t)^2.
\end{equation}

Therefore the metric takes the form
\begin{equation}\label{metric:2}
  \mathrm{d}s^2 = -\mathrm d t^2 +a(t)^2\left(e^{2\beta_1(t)}\mathrm dx^2+ e^{2\beta_2(t)}\mathrm dy^2 + e^{2\beta_3(t)}\mathrm dz^2\right).
\end{equation}
The metric of the form \eqref{metric:2} has been introduced in the paper\cite{kumar}.
The velocity vector $\dot \beta(t)$ has norm $\sigma(t)$ and therefore can be written as
$\dot \beta(t) = \sigma(t) \hat \beta(t)$, where $\hat \beta(t)$ is a unit vector for  all $t$. Assuming that $\beta(t)$ lies in a plain $z=0$ one obtains
\begin{equation}
  \dot \beta(t) = \sigma(t) (\cos \theta(t), \sin \theta(t),0),
\end{equation}
for some function $\theta(t)$ which will be determined later. Direct integration yields that
\begin{equation}
\beta(t) =  \left(\int \sigma(t) \cos \theta(t)\mathrm d t,\int \sigma(t) \sin \theta(t) \mathrm d t,0\right).
\end{equation}

Since we have a curve $\beta(t)$ lying in plane $z=0$ and we need it to be in plane $x+y+z=0$,
 it remains to find the rotation of the space that maps plane $z=0$ to $x+y+z=0$. Let us recall that each rotation can be written as a composition of three rotations around coordinate axis using Euler angles. Therefore arbitrary rotation in space can be expressed with the following matrix
\begin{equation}
M = \left(
\begin{array}{ccc}
 \cos \zeta  \cos \eta  \cos \xi -\sin \zeta \sin \xi\hspace{2.5mm} &
 -\cos \eta \cos \xi \sin \zeta -\cos \zeta \sin \xi\hspace{2.5mm}  &
 \cos \xi \sin \eta \\
 \cos \xi  \sin \zeta +\cos \zeta \cos \eta \sin \xi\hspace{2.5mm}  &
 \cos \zeta \cos \xi -\cos \eta  \sin \zeta \sin \xi\hspace{2.5mm}  &
 \sin \eta  \sin \xi \\
 -\cos \zeta \sin \eta  \hspace{2.5mm}&
  \sin \zeta \sin \eta  \hspace{2.5mm}&
  \cos \eta \\
\end{array}
\right).
\end{equation}

Since $M$ is orthogonal matrix, which preserves lengths and angles it is sufficient to map normal vector to normal vector
\begin{equation}
  \frac 1{\sqrt 3} \left(
                     \begin{array}{c}
                       1 \\
                       1 \\
                       1 \\
                     \end{array}
                   \right) = M \left(
                                 \begin{array}{c}
                                   0 \\
                                   0 \\
                                   1 \\
                                 \end{array}
                               \right).
\end{equation}
The solution is given by
\begin{equation}
M = \left(
\begin{array}{ccc}
 \frac{\cos \zeta}{\sqrt{6}}-\frac{\sin \zeta }{\sqrt{2}} \hspace{2.5mm}&
-\frac{\cos \zeta}{\sqrt{2}}-\frac{\sin \zeta }{\sqrt{6}} \hspace{2.5mm}&
 \frac{1}{\sqrt{3}} \\
 \frac{\cos \zeta}{\sqrt{6}}+\frac{\sin \zeta}{\sqrt{2}} \hspace{2.5mm}&
 \frac{\cos \zeta}{\sqrt{2}}-\frac{\sin \zeta}{\sqrt{6}} \hspace{2.5mm}&
 \frac{1}{\sqrt{3}}\\
 -\sqrt{\frac{2}{3}} \cos \zeta  \hspace{2.5mm}&
  \sqrt{\frac{2}{3}} \sin \zeta  \hspace{2.5mm}&
   \frac{1}{\sqrt{3}} \\
\end{array}
\right).
\end{equation}
For the purpose of the sequel it is sufficient to take one solution, therefore let $\zeta=0$ and
\begin{equation}
M = \left(
\begin{array}{ccc}
 \frac{1}{\sqrt{6}} &
-\frac{1}{\sqrt{2}} &
 \frac{1}{\sqrt{3}} \\
 \frac{1}{\sqrt{6}} &
 \frac{1}{\sqrt{2}} &
 \frac{1}{\sqrt{3}}\\
 -\sqrt{\frac{2}{3}} &
  0 &
   \frac{1}{\sqrt{3}} \\
\end{array}
\right).
\end{equation}

The curve $\beta(t)$ then takes the form
\begin{align}\label{beta}
\beta(t)&= M \left(\begin{array}{c}
              \int \sigma (t) \cos \theta(t) \mathrm d t \\
              \int \sigma (t) \sin \theta(t) \mathrm d t \\
              0
            \end{array}\right)
        = \left(\begin{array}{c}
            \frac{\int \sigma (t) \cos \theta (t) \,dt}{\sqrt{6}} - \frac{\int \sigma (t) \sin \theta (t) \, dt}{\sqrt{2}} \\
            \frac{\int \sigma(t) \sin \theta (t) \, dt}{\sqrt{2}}+\frac{\int \sigma (t) \cos \theta (t) \, dt}{\sqrt{6}}\\
            -\sqrt{\frac{2}{3}} \int \sigma (t) \cos \theta (t) \, dt\end{array}\right).
\end{align}
\section{Model and EOM}
In this paper we discuss the action given by
\begin{align} \label{action:1}
S = \frac{1}{16 \pi G}  \int \big(R- 2 \Lambda  + P(R){\FF}(\Box) Q(R)\big) \sqrt{-g}\, \mathrm d^4 x ,
\end{align}
where the Universe is represented by a pseudo-Riemannian manifold with metric $g_{\mu\nu}$ of signature $(1,3)$, $P$ and $Q$ are differentiable functions of scalar curvature $R$, $\Lambda$ is cosmological constant and $\FF(\Box) = \sum_{n=1}^{+\infty} f_n \Box^n + \sum_{n=1}^{+\infty} f_{-n} \Box^{-n}$ is nonlocal operator. It is obvious that this model includes GR if we set $\FF(\Box) =0$. Since the emphasis in this paper is on the nonlocal modification of the gravity we will not include  matter term in the action.
The first step is derivation of equations of motion, which is a lengthy procedure that was presented in \cite{biswas4}, in particular\cite{dimitrijevic9},
\begin{align}\label{eom:1}
  &G_{\mu\nu} + \Lambda g_{\mu\nu} - \frac 12 g_{\mu\nu} P \FF(\Box) Q+ R_{\mu\nu} W - K_{\mu\nu} W + \frac 12 \Omega_{\mu\nu} =0,
\end{align}
where
\begin{align}
  W & = P'(R) \FF(\Box) Q(R) + Q'(R)\FF(\Box)P(R),\\
  K_{\mu\nu} &= \nabla_\mu \nabla_\nu - g_{\mu\nu}\Box, \\
  S_{\mu\nu}(A,B) &= g_{\mu\nu} \nabla_\lambda A \nabla^{\lambda} B + g_{\mu\nu}A \Box B -2 \nabla_\mu A \nabla_\nu B, \\
  \Omega_{\mu\nu} &= \sum_{n=1}^{+\infty}f_n \sum_{l=0}^{n-1} S_{\mu\nu}(\Box^l P,\Box^{n-1-l}Q)  \nonumber \\
   &- \sum_{n=1}^{+\infty}f_{-n} \sum_{l=0}^{n-1} S_{\mu\nu}(\Box^{-(l+1)} P,\Box^{-(n-l)}Q),
\end{align}
and $'$ denote derivation over $R$.
Equation \eqref{eom:1} contains infinitely many derivatives and therefore we cannot find its general solution, but we can simplify it considerably by choosing $P=Q$ and moreover we let $Q$ be an eigenfunction of the operator $\Box$ with eigenvalue $q$. Hence, equation \eqref{eom:1} is simplified
\begin{align}\label{eom:2}
  &G_{\mu\nu} + \Lambda g_{\mu\nu} - \frac 12 g_{\mu\nu}  \FF(q) Q^2+ R_{\mu\nu} W - K_{\mu\nu} W + \frac 12 \Omega_{\mu\nu} =0,
\end{align}
where
\begin{align}
  W & = 2 Q'(R) \FF(\Box) Q(R),\\
  \Omega_{\mu\nu} &= \FF'(q) S_{\mu\nu}(Q,Q).
\end{align}
This equation transforms as
\begin{align}
  &(G_{\mu\nu} + \Lambda g_{\mu\nu})(1+ 2\FF(q) Q Q') + \FF(q)g_{\mu\nu}(-\frac 12 Q^2 + QQ' (R-2\Lambda)) \nonumber \\
  & -2 \FF(q)K_{\mu\nu} QQ'+ \frac 12 \FF'(q) S_{\mu\nu}(Q,Q) =0.
\end{align}

In particular, the most interesting case for us is $Q = \sqrt{R-2\Lambda}$, which gives us $Q Q' = \frac 12$ and equations of motion get the form
\begin{align}\label{eom:3}
  &(G_{\mu\nu} + \Lambda g_{\mu\nu})(1+ \FF(q)) + \frac 12 \FF'(q) S_{\mu\nu}(Q,Q) =0.
\end{align}
It is clear that if we choose function $\FF$ such that
\begin{align}\label{fbox:1}
  \FF(q) = -1, \qquad \FF'(q) = 0,
\end{align}
then equation \eqref{eom:3} is satisfied. Therefore the next section is devoted to solving the following eigenvalue problem
\begin{equation}\label{evp:1}
  \Box \sqrt{R-2\Lambda} = q \sqrt{R-2\Lambda}.
\end{equation}
From the previous discussion we conclude that if we solve \eqref{evp:1} and function $\FF$ is constrained by \eqref{fbox:1} we know that equations of motion \eqref{eom:1} are satisfied as well.

\section{Cosmological solutions}
In the beginning, it is interesting to note that Ricci tensor, scalar curvature and d'Alembertian of the metric \eqref{metric:2} do not depend on  $\theta(t)$ and hence it will remain undetermined in the following calculations.
\begin{align}
  R &= R_{FLRW} + \sigma^2, \\
  \Box u(t)&= \Box_{FLRW}u(t), \\
  R_{00} &= R_{00,FLRW} - \sigma^2,\\
  G_{00} &= G_{00,FLRW} - \frac 12\sigma^2,
\end{align}
where index $FLRW$ denotes quantities corresponding to the $FLRW$ metric with scale factor $a(t)$ and $k=0$.
\subsection{Scale factor in exponential form $a(t)=e^{\gamma t^2}$}
As a first case we take
\begin{equation}
  a(t)= A e^{\gamma t^2},
\end{equation}
then the eigenvalue problem \eqref{evp:1} takes the form
\begin{align}
  &4 \sigma(t)^2 \left(q \left(6 \gamma-\Lambda +24 \gamma ^2t^2\right)+12 \gamma ^2 \left(6
   \gamma  t^2+1\right)\right)+4 q \left(-6 \gamma +\Lambda -24\gamma ^2 t^2\right)^2 \nonumber \\
   &+q \sigma(t)^4 + 2 \left(6 \gamma -\Lambda+24 \gamma ^2 t^2\right) \dot \sigma(t)^2+2 \sigma (t) \Big(\left(6 \gamma -\Lambda +24 \gamma ^2 t^2\right) \ddot \sigma(t)\nonumber \\
   &+6 \gamma t \left(-2 \gamma -\Lambda +24 \gamma ^2 t^2\right) \dot \sigma(t)\Big)+96 \gamma ^2 \left(-\Lambda +144 \gamma ^3
   t^4+36 \gamma ^2 t^2+\gamma \left(6-6 \Lambda t^2\right)\right)\nonumber \\
   &+\sigma (t)^3 \left(6 \gamma  t \dot \sigma(t)+\ddot\sigma(t)\right) = 0.
\end{align}
The  remaining equation in $\sigma(t)$ is nonlinear second order equation and we can obtain only particular solutions. To simplify further we take $\gamma = \frac\Lambda6$ and $q=-\Lambda$ as we have in the corresponding $FLRW$ solution (see \cite{dimitrijevic11} for details) one obtains the following equation in $\sigma(t)$
\begin{align}
  &4 \Lambda ^2 t^2 \dot\sigma(t)^2+4 \Lambda ^2 t \sigma (t) \left(\left(\Lambda  t^2-2\right)
   \dot \sigma(t)+t \ddot\sigma(t)\right)-4 \Lambda ^2 \left(\Lambda  t^2-1\right) \sigma
   (t)^2\nonumber \\
   &+3 \sigma (t)^3 \left(\Lambda t \dot \sigma(t)+\ddot \sigma(t)\right)-3 \Lambda  \sigma(t)^4 =0.
\end{align}
One particular solution is
\begin{equation}
  \sigma(t) =\sigma_0 t,
\end{equation}
for some arbitrary constant $\sigma_1$.
Hence we obtained the solution of eigenvalue problem \eqref{evp:1} in the form
\begin{equation}
  q = -\Lambda, \qquad a(t) = A e^{\frac\Lambda6 t^2}, \qquad \sigma(t) =\sigma_0 t,
\end{equation}
which is also a solution of EOM if
\begin{equation}
  \FF(\frac \Lambda6) = -1, \qquad \FF'(\frac \Lambda6) = 0,
\end{equation}
as we have already seen in the previous section.
\subsection{Scale factor as a linear combination of exponential functions $a(t) = \alpha e^{\lambda t} + \beta e^{-\lambda t}$}
As a second case let us take
\begin{equation}
    a(t) = \alpha e^{\lambda t} + \beta e^{-\lambda t} \mbox{ and }\sigma(t) = \sigma_0 a(t)^{-2}.
\end{equation}
Inserting these values into \eqref{evp:1} yields an equation of the form
\begin{equation}
  \sum_{n=0}^{8} A_n e^{2\lambda n t} =0,
\end{equation}
where the coefficients $A_n$ are given by
\begin{align}
  A_0&=4 \beta ^8 q \left(6 \lambda^2-\Lambda \right)^2,\\
  A_1&=16 \alpha \beta ^7 \left(6 \lambda^2-\Lambda \right) \left(3 \lambda ^4+6 \lambda ^2 q-2 \Lambda q\right), \\
  A_2&=4 \beta ^4 \Big(144\alpha ^2 \beta ^2 \lambda ^6-72 \alpha ^2 \beta ^2 \lambda ^4
   \Lambda +288 \alpha ^2 \beta ^2\lambda ^4 q-192 \alpha ^2 \beta^2 \lambda ^2 \Lambda  q \nonumber \\
   &+28 \alpha^2 \beta ^2 \Lambda ^2 q+6 \lambda^2 q \sigma_0^2-\Lambda  q
   \sigma_0^2+6 \lambda ^4\sigma_0^2-\lambda ^2 \Lambda\sigma_0^2\Big), \\
   A_3&=8 \alpha  \beta^3 \Big(252 \alpha ^2 \beta ^2 \lambda ^6-90 \alpha ^2 \beta ^2\lambda ^4 \Lambda +216 \alpha ^2 \beta ^2 \lambda ^4 q-156 \alpha^2 \beta ^2 \lambda ^2 \Lambda q \nonumber \\
   &+28 \alpha ^2 \beta ^2 \Lambda ^2 q+6 \lambda ^2 q \sigma_0^2-2
   \Lambda  q \sigma_0^2-3 \lambda ^4 \sigma_0^2+2 \lambda ^2
   \Lambda  \sigma_0^2\Big), \\
   A_4&=3456 \alpha ^4 \beta ^4 \lambda ^6-960 \alpha ^4 \beta ^4 \lambda ^4
   \Lambda +2016 \alpha ^4 \beta ^4 \lambda ^4 q-1440 \alpha ^4 \beta
   ^4 \lambda ^2 \Lambda  q \nonumber \\
   &+280 \alpha ^4 \beta ^4 \Lambda ^2 q+q s_0^4+48 \alpha ^2 \beta ^2 \lambda ^2 q \sigma_0^2-24 \alpha ^2 \beta ^2 \Lambda  q \sigma_0^2-2 \lambda ^2
   \sigma_0^4 \nonumber \\
   &-96 \alpha ^2 \beta ^2 \lambda ^4 \sigma_0^2+40 \alpha
   ^2 \beta ^2 \lambda ^2 \Lambda \sigma_0^2, \\
   A_5&=8 \alpha ^3 \beta \Big(252 \alpha ^2 \beta ^2 \lambda ^6-90 \alpha ^2 \beta ^2 \lambda ^4 \Lambda +216 \alpha ^2 \beta ^2 \lambda ^4 q-156 \alpha
   ^2 \beta ^2 \lambda ^2 \Lambda q \nonumber \\
   &+28 \alpha ^2 \beta ^2 \Lambda ^2 q+6 \lambda ^2 q \sigma_0^2-2 \Lambda  q \sigma_0^2-3 \lambda^4 \sigma_0^2+2 \lambda ^2
   \Lambda  \sigma_0^2\Big), \\
   A_6&=4\alpha ^4 \Big(144 \alpha ^2\beta ^2 \lambda ^6-72 \alpha ^2\beta ^2 \lambda ^4 \Lambda +288 \alpha ^2 \beta ^2 \lambda ^4q -192 \alpha ^2 \beta ^2 \lambda^2 \Lambda  q \nonumber \\
   &+28 \alpha ^2 \beta ^2 \Lambda ^2 q+6 \lambda ^2 q\sigma_0^2-\Lambda  q \sigma_0^2+6 \lambda ^4\sigma_0^2-\lambda ^2 \Lambda
   \sigma_0^2\Big), \\
   A_7&=16 \alpha ^7 \beta  \left(6 \lambda ^2-\Lambda\right) \left(3 \lambda ^4+6\lambda ^2 q-2 \Lambda q\right), \\
   A_8&=4\alpha ^8 q \left(6 \lambda^2-\Lambda \right)^2.
\end{align}
Thus, we need to solve the system
\begin{equation}
  A_n=0,\qquad n=\overline{0,8}.
\end{equation}
The highest order coefficient $A_8$ vanish if $\lambda^2 = \frac\Lambda6$, while the remaining coefficients are simplified to

\begin{align}
  A_2&=576 \alpha ^2 \beta ^6 \lambda ^4 \left(q-2 \lambda
   ^2\right),\\
  A_3&=24 \alpha  \beta ^3 \lambda ^2 \left(-96 \alpha ^2\beta ^2 \lambda ^4+96 \alpha ^2\beta ^2 \lambda ^2 q-2 q
   \sigma_0^2+3 \lambda ^2\sigma_0^2\right), \\
  A_4&=-2304 \alpha ^4\beta ^4 \lambda ^6+3456 \alpha ^4 \beta ^4 \lambda ^4 q +q\sigma_0^4, \\
  A_5&=-96 \alpha ^2 \beta ^2
   \lambda ^2 q \sigma_0^2-2 \lambda^2 \sigma_0^4+144 \alpha ^2 \beta^2 \lambda ^4 \sigma_0^2, \\
  A_6&=24 \alpha ^3 \beta  \lambda ^2\left(-96 \alpha ^2 \beta ^2 \lambda ^4+96 \alpha ^2 \beta ^2\lambda ^2 q-2 q \sigma_0^2+3
   \lambda ^2 \sigma_0^2\right), \\
  A_7&=576\alpha ^6 \beta ^2 \lambda ^4 \left(q-2 \lambda ^2\right).
\end{align}
Now we set $q = 2\lambda^2$ and the remaining equations are
\begin{align}
  24 \alpha  \beta ^3\lambda ^4 \left(96 \alpha ^2\beta ^2 \lambda^2-\sigma_0^2\right) =0, \\
  -48 \alpha^2  \beta ^2\lambda ^4 \left(96 \alpha ^2\beta ^2 \lambda^2-\sigma_0^2\right) =0, \\
  24 \alpha^3  \beta\lambda ^4 \left(96 \alpha ^2\beta ^2 \lambda^2-\sigma_0^2\right) =0.
\end{align}
Hence we get a solution
\begin{equation}
  a(t)= \alpha e^{\lambda t} + \beta e^{-\lambda t} \mbox{ and }\sigma(t) = \sigma_0 \left(\alpha e^{\lambda t} + \beta e^{-\lambda t}\right)^{-2},
\end{equation}
in the following  two  cases

\begin{enumerate}
 \item $\lambda = \pm \sqrt{\frac\Lambda6}$, $\quad q = \frac \Lambda3$,
 $\quad \sigma_0^2 = 16 \alpha^2 \beta^2 \Lambda$, \\
 \item $\lambda = \pm \sqrt{\frac\Lambda6}$, $\quad q = \frac \Lambda3$, $\quad\alpha \beta=0$.
\end{enumerate}

For example, in the first case we get scale factors of the form $a(t)= A \cosh \lambda t$ and $a(t)= A \sinh \lambda t$, while in the second case we get $a(t)= \alpha e^{\lambda t}$.
\subsection{Scale factor $a(t)= \left(\alpha e^{\lambda t} + \beta e^{-\lambda t}\right)^{\frac 12}$}
Also one can take scale factor $a(t)$ in the form
\begin{equation}
  a(t)= \left(\alpha e^{\lambda t} + \beta e^{-\lambda t}\right)^{\frac 12},
\end{equation}
and the condition \eqref{evp:1} is transformed into
\begin{align}
  &2 \left(3 \lambda ^2-2 \Lambda
   \right) \left(\beta +\alpha  e^{2
   \lambda  t}\right) \left(3 \lambda
   ^2 q-2 \Lambda  q+ \dot \sigma(t)^2\right) \nonumber \\&+4 q \left(3 \lambda
   ^2-2 \Lambda \right) \sigma (t)^2
   \left(\beta +\alpha  e^{2 \lambda
   t}\right)+2 q \sigma (t)^4
   \left(\beta +\alpha  e^{2 \lambda
   t}\right)\nonumber \\
   &+\left(3 \lambda ^2-2 \Lambda \right) \sigma (t) \left(2
   \ddot\sigma (t) \left(\beta +\alpha
   e^{2 \lambda  t}\right)+3 \lambda
   \dot \sigma(t) \left(\alpha  e^{2
   \lambda  t}-\beta
   \right)\right)\nonumber \\
   &+\sigma (t)^3
   \left(2 \ddot \sigma(t) \left(\beta
   +\alpha  e^{2 \lambda  t}\right)+3
   \lambda  \dot \sigma(t) \left(\alpha
   e^{2 \lambda  t}-\beta
   \right)\right) = 0.
\end{align}
This expression is simplified by taking $\lambda^2 = \frac 23 \Lambda$
\begin{align}
\sigma (t)^3 \left(2 q \sigma (t) \left(\beta +\alpha  e^{2 \lambda
   t}\right)+2 \ddot\sigma(t) \left(\beta +\alpha  e^{2 \lambda
   t}\right)+3 \lambda  \dot \sigma(t) \left(\alpha  e^{2 \lambda
   t}-\beta \right)\right) = 0.
\end{align}
The last equation has obvious solution $\sigma(t)=0$ and its general solution is expressed in terms of hypergeometric functions
\begin{align}
  \sigma(t) &= C_1  \left(e^{\lambda t}\sqrt{\frac\alpha\beta}\right)^{\frac 34 -\eta} \;_2F_1\left(\frac{3}{4}, \frac{3}{4}-\eta;1-\eta; -\frac{\alpha e^{2 \lambda  t}}{\beta }\right) \nonumber \\
  &+ C_2  \left(e^{\lambda t}\sqrt{\frac\alpha\beta}\right)^{\frac 34 +\eta} \;_2F_1\left(\frac{3}{4}, \frac{3}{4}+\eta;1+\eta; -\frac{\alpha e^{2 \lambda  t}}{\beta }\right),
\end{align}
where $\eta = \sqrt{\frac 9{16} - \lambda^{-2} q}$. As an example one can take $\eta = \frac 12$ and hence $q = \frac 5{16}\lambda^2$ which gives us
\begin{align}
  \sigma(t) &= \left(e^{\lambda t}\sqrt{\frac\alpha\beta}\right)^{\frac 14}
  \left(C_1
  \frac {\sqrt{1+\sqrt{1+\frac\alpha\beta e^{2\lambda t}}}}{\sqrt{1 +  \frac\alpha\beta e^{2\lambda  t}}}
  + C_2 \frac{\sin \left(\frac 12 \arctan \sqrt{\frac\alpha\beta}e^{\lambda t} \right)}{\sqrt[4]{1 +  \frac\alpha\beta e^{2\lambda  t}}}
  \right).
\end{align}

\subsection{Constant $\sigma$ solutions}
Consider the scale factor
\begin{align*}
  a(t) &= t^n(\alpha e^{\gamma t^2} + \beta e^{-\gamma t^2}),\\
  \sigma(t) &= \sigma_0 = \mathrm{const},
\end{align*}
and the eigenvalue problem is transformed into polynomial equation over $e^{2\gamma t^2}$
\begin{equation}
  \sum_{j=0}^6 B_j e^{2j \gamma t^2}=0.
\end{equation}
From the highest order term we get $n(2n-1)(3n-2)=0$.
On the other hand we see that $\alpha\beta =0$ and without loss of generality we set $\beta =0$.
In case $n=\frac 23$ it is remaining to solve
\begin{align}
&\frac{4}{9} \alpha ^6 \left(108 \gamma -6 \Lambda +4 q+3
   \sigma_0^2\right) =0, \\
&\frac{8}{3} \alpha ^6 \left(12 \gamma ^2+6 \gamma
   \Lambda -3 \gamma  \sigma_0^2+44 \gamma  q-2 \Lambda  q+q
   \text{$\sigma $0}^2\right) =0,\\
& \alpha^6 \big(6336 \gamma ^3-288 \gamma ^2 \Lambda +144 \gamma ^2
    \sigma_0^2+2064 \gamma ^2 q \nonumber \\
&-176 \gamma  \Lambda  q+88 \gamma q  \sigma_0^2+4 \Lambda
   ^2 q-4 \Lambda  q \sigma_0^2+q  \sigma_0^4\big) =0 ,\\
& 96 \alpha ^6 \gamma^2 \left(180 \gamma ^2-6 \gamma
   \Lambda +3 \gamma \sigma_0^2+44 \gamma  q-2 \Lambda q+q\sigma_0^2\right)=0, \\
&2304 \alpha ^6 \gamma ^4 (6 \gamma+q)=0.
\end{align}
It is evident that $q=-6\gamma$ and after substitution one finds that
\begin{equation}
 \sigma_0^2=2\Lambda-28\gamma.
\end{equation}
Instead of $\sigma_0$ we will introduce parameter $\eta$ such that $\sigma_0^2 = 2\Lambda \eta$ and hence the final solution reads
\begin{align}
  a(t) &= A t^{2/3}e^{\frac\Lambda{14}(1-\eta) t^2},\label{sol:1} \\
  q &= -\frac 37 \Lambda (1-\eta), \\
  \sigma^2 &= 2\Lambda \eta.
\end{align}

In case $n=\frac 12$ we conclude that $\beta =0$ and $q=-6\gamma$ in the same way as in the previous case. Hence we have the following conditions
\begin{align}
  &-6 \alpha ^6 \gamma  \left(16
   \gamma -2 \Lambda +\sigma_0^2\right) \left(36 \gamma -2
   \Lambda +\sigma_0^2\right) =0, \\
   &-288 \alpha ^6 \gamma^3 \left(24 \gamma -2 \Lambda+\sigma_0^2\right)=0,
   \end{align}
which clearly has no solution.

Finally the third case $n=0$ was discussed previously.
The solution \eqref{sol:1} converges to an isotropic solution as $\eta$ tends to $0$. This isotropic solution have been found  and discussed in papers\cite{dimitrijevic11,dimitrijevic12}. Moreover there are several more solutions of the flat $FLRW$ model that can be extended to the anisotropic case with constant $\sigma$. The other solutions of the $FLRW$ model can be treated similarly and provide the following solutions

\begin{align*}
  a_1(t) &= A \cosh^{\frac{2}{3}}(\sqrt{\frac{3 \Lambda}{8}} (1-\eta) \; t) \; , \quad
  q = \frac{3 \Lambda}{8} (1-\eta)^{2}, \qquad
  \sigma^{2} = 2 \Lambda \eta (2-\eta),
\end{align*}

\begin{align*}
  a_2(t) &= A \sinh^{\frac{2}{3}}(\sqrt{\frac{3 \Lambda}{8}} (1-\eta) \; t) \; , \quad
  q = \frac{3 \Lambda}{8} (1-\eta)^{2}, \qquad
  \sigma^{2} = 2 \Lambda \eta (2-\eta),
\end{align*}

\begin{align*}
  a_3(t) &= A \cos^{\frac{2}{3}}(\sqrt{-\frac{3 \Lambda}{8}} (1-\eta) \; t) \; , \quad
  q = \frac{3 \Lambda}{8} (1-\eta)^{2}, \qquad
  \sigma^{2} = 2 \Lambda \eta (2-\eta),
\end{align*}
\begin{align*}
  a_4(t) &= A \sin^{\frac{2}{3}}(\sqrt{-\frac{3 \Lambda}{8}} (1-\eta) \; t) \; , \quad
  q = \frac{3 \Lambda}{8} (1-\eta)^{2}, \qquad
  \sigma^{2} = 2 \Lambda \eta (2-\eta).
\end{align*}
It is worth noting that in case of solutions $a_1(t)$ and $a_2(t)$ cosmological constant $\Lambda$ is positive, while for $a_3(t)$ and $a_4(t)$ cosmological constant $\Lambda$ is negative.

\subsection{$FLRW$ solutions as anisotropic solutions}
The expressions for scalar curvature for the metric \eqref{metric:2} and $FLRW$ metric (with arbitrary $k$) are
\begin{align}
  R &= \frac{6 \left(a(t) \ddot a(t)+\dot a(t)^2\right)}{a(t)^2}+\sigma (t)^2, \label{scur:1}\\
  R_{FLRW} &= \frac{6 \left(a(t) \ddot a(t)+\dot a(t)^2 +k \right)}{a(t)^2}.\label{scur:2}
\end{align}
Comparing these expressions for scalar curvature we can see that if we choose scale factor $a(t)$ such that it is a solution of $FLRW$ model with $k\neq 0$ and $\sigma = \sigma_0 a(t)^{-1}$ we see that expressions \eqref{scur:1} and \eqref{scur:2} are equal if we set $\sigma_0^2 = 6k $. Therefore each scale factor $a(t)$ which is a solution of the $FLRW$ model with $k=1$ can be extended to  an anisotropic solution by the formulas \eqref{ai} and \eqref{beta}. According to the paper \cite{dimitrijevic13} there are three such scale factors have been found
\begin{itemize}
  \item $a(t) = A e^{\pm \sqrt {\frac 16 \Lambda} \; t}$, $\sigma(t) = \frac{\sqrt 6}A e^{\mp \sqrt {\frac 16 \Lambda} \; t}$,
  \item $a(t) = A \cosh^{\frac 12} {\sqrt {\frac 23\Lambda} \; t}$, $\sigma(t) = \frac{\sqrt 6}A \cosh^{-\frac 12} {\sqrt {\frac 23\Lambda} \; t}$,
  \item $a(t) = A \sinh^{\frac 12} {\sqrt {\frac 23\Lambda} \; t}$, $\sigma(t) = \frac{\sqrt 6}A \cosh^{-\frac 12} {\sqrt {\frac 23\Lambda} \; t}$.
\end{itemize}

On the other hand if we take any $FLRW$ solution for $k=0$ in nonlocal de Sitter model \eqref{eq.1.1} and choose $\sigma(t)$ in the following way
\begin{align}
  \sigma(t)^2 &=\sigma_0^2\left(\frac{6 \left(a(t) \ddot a(t)+\dot a(t)^2\right)}{a(t)^2}-2 \Lambda\right),
\end{align}
we see that $R-2\Lambda$ and $R_{FLRW} -2\Lambda$ are proportional, therefore each solution of $FLRW$ model is an anisotropic solution as well with this choice of $\sigma(t)$.
There are (at least) eight such solutions
\begin{itemize}
  \item $a(t) = A t^{\frac 23} e^{\frac\Lambda{14} t^2}$, $\quad\sigma(t)= \sigma_0 t^{-1}(7+3\Lambda t^2)$,
  \item $a(t) = A  e^{\frac\Lambda6 t^2}$, $\qquad\sigma(t) = \sigma_0 \Lambda t$,
  \item $a(t) = A \cosh^{\frac 23} \sqrt{\frac 38 \Lambda}\,t$, $\quad\sigma(t) = \sigma_0 \sqrt{10\Lambda -9\Lambda
   \cosh^{-2}\sqrt{\frac 38 \Lambda } t}$,
  \item $a(t) = A \sinh^{\frac 23} \sqrt{\frac 38 \Lambda}\,t$, $\quad\sigma(t) = \sigma_0 \sqrt{10\Lambda +9 \Lambda
   \sinh^{-2}\sqrt{\frac 38 \Lambda } t}$,
  \item $a(t) = A \left(1\pm \sin \sqrt{-\frac 32 \Lambda}\,t\right)^{\frac 13}$, $\quad\sigma(t) = \frac{\sigma_0}{\sqrt{\pm 1 +\sin \sqrt{-\frac 32 \Lambda} t}}$,
  \item $a(t) = A \cos^{\frac 23} \sqrt{-\frac 38 \Lambda}\,t$, $\quad\sigma(t) = \frac{\sigma_0}{\sqrt{1+\cos \sqrt{-\frac 32 \Lambda}\,t}}$,
  \item $a(t) = A \sin^{\frac 23} \sqrt{-\frac 38 \Lambda}\,t$, $\quad\sigma(t) = \frac{\sigma_0}{\sqrt{-1+\cos \sqrt{-\frac 32 \Lambda}\,t}}$.
\end{itemize}
\section{Concluding Remarks}
In this paper, several anisotropic and homogeneous Bianchi I cosmological solutions of nonlocal de Sitter gravity model \eqref{eq.1.1} are presented. Anisotropy depends on two time dependent parameters $\sigma (t)$ and $\theta(t)$. Equations of motion contain $\sigma (t)$, while $\theta (t)$ remains undetermined. These anisotropic solutions are an extension of the corresponding homogeneous and isotropic ones, and when parameter $\sigma (t)$ tends to zero, anisotropy disappears.

It is worth noting that anisotropic cosmological solutions may be important not only for research of space-time dynamics at early stage of the Universe but also at late  cosmic scales in the framework of dipole cosmology\cite{krishnan} with evidence of some dipole anisotropy, what may have implications for the currently debated cosmic tensions (including $H_0$).

 Simple nonlocal de Sitter gravity model \eqref{eq.1.1} shows rich spectrum of so far obtained cosmological solutions, see also\cite{dimitrijevic10,dimitrijevic13}. We plan to continue with exploration of new cosmological possibilities of \eqref{eq.1.1}.

%
%
%
%

\section*{Acknowledgments}

I would like to thank Branko Dragovich, Zoran Rakic and Jelena Stankovic for numerous discussions and comments about the paper.

This research was partially funded by the Ministry of Education, Science and Technological Developments of the Republic of Serbia: grant number 451-03-47/2023-01/ 200104 with University of Belgrade, Faculty of Mathematics. It is also
partially supported by the COST Action: CA21136 – Addressing observational tensions in cosmology with
systematics and fundamental physics (CosmoVerse).

%
%


\end{document}